\documentclass[aps,prl,twocolumn,superscriptaddress]{revtex4-1}
\usepackage{graphicx}

\newcommand{\hebar}{\overline{\mathrm{He}}}

\begin{document}

\title{Search for Antihelium with the BESS-Polar Spectrometer}

\author{K.\thinspace Abe}
\thanks{Present address: ICRR, The University of Tokyo, Kamioka, Gifu, 506-1205, Japan}
\affiliation{Kobe University, Kobe, Hyogo 657-8501, Japan.}
\author{H.\thinspace Fuke}
\affiliation{Institute of Space and Astronautical Science, Japan Aerospace Exploration Agency (ISAS/JAXA), Sagamihara, Kanagawa 252-5210, Japan.}
\author{S.\thinspace Haino}
\thanks{Present address: Instituto Nazionale di Fisica Nucleare (INFN), Perugia, Italy}
\affiliation{High Energy Accelerator Research Organization (KEK), Tsukuba, Ibaraki 305-0801, Japan.}
\author{T.\thinspace Hams}
\thanks{Also University of Maryland, Baltimore County, Baltimore, MD 21250, USA}
\affiliation{NASA/Goddard Space Flight Center, Greenbelt, MD 20771, USA.}
\author{M.\thinspace Hasegawa}
\affiliation{High Energy Accelerator Research Organization (KEK), Tsukuba, Ibaraki 305-0801, Japan.}
\author{A.\thinspace Horikoshi}
\affiliation{High Energy Accelerator Research Organization (KEK), Tsukuba, Ibaraki 305-0801, Japan.}
\author{A.\thinspace Itazaki}
\affiliation{Kobe University, Kobe, Hyogo 657-8501, Japan.}
\author{K.\thinspace C.\thinspace Kim}
\affiliation{University of Maryland, College Park, MD 20742, USA.}
\author{T.\thinspace Kumazawa}
\affiliation{High Energy Accelerator Research Organization (KEK), Tsukuba, Ibaraki 305-0801, Japan.}
\author{A.\thinspace Kusumoto}
\affiliation{Kobe University, Kobe, Hyogo 657-8501, Japan.}
\author{M.\thinspace H.\thinspace Lee}
\affiliation{University of Maryland, College Park, MD 20742, USA.}
\author{Y.\thinspace Makida}
\affiliation{High Energy Accelerator Research Organization (KEK), Tsukuba, Ibaraki 305-0801, Japan.}
\author{S.\thinspace Matsuda}
\affiliation{High Energy Accelerator Research Organization (KEK), Tsukuba, Ibaraki 305-0801, Japan.}
\author{Y.\thinspace Matsukawa}
\affiliation{Kobe University, Kobe, Hyogo 657-8501, Japan.}
\author{K.\thinspace Matsumoto}
\affiliation{High Energy Accelerator Research Organization (KEK), Tsukuba, Ibaraki 305-0801, Japan.}
\author{J.\thinspace W.\thinspace Mitchell}
\affiliation{NASA/Goddard Space Flight Center, Greenbelt, MD 20771, USA.}
\author{Z.\thinspace Myers}
\thanks{Present address: Physics Department, Technion – Israel Institute of Technology, Technion City, Haifa, Israel}
\affiliation{University of Maryland, College Park, MD 20742, USA.}
\author{J.\thinspace Nishimura}
\affiliation{The University of Tokyo, Bunkyo, Tokyo 113-0033 Japan.}
\author{M.\thinspace Nozaki}
\affiliation{High Energy Accelerator Research Organization (KEK), Tsukuba, Ibaraki 305-0801, Japan.}
\author{R.\thinspace Orito}
\thanks{Present address: Tokushima University, Tokushima, Japan}
\affiliation{Kobe University, Kobe, Hyogo 657-8501, Japan.}
\author{J.\thinspace F.\thinspace Ormes}
\affiliation{University of Denver, Denver, CO 80208, USA.}
\author{K.\thinspace Sakai}
\thanks{Present address: University of Maryland, Baltimore County, Baltimore, MD 21250, USA}
\affiliation{The University of Tokyo, Bunkyo, Tokyo 113-0033 Japan.}
\author{M.\thinspace Sasaki}
\email[Corresponding author. Email: ]{Makoto.Sasaki@nasa.gov}
\thanks{Also University of Maryland, College Park, MD 20742, USA}
\affiliation{NASA/Goddard Space Flight Center, Greenbelt, MD 20771, USA.}
\author{E.\thinspace S.\thinspace Seo}
\affiliation{University of Maryland, College Park, MD 20742, USA.}
\author{Y.\thinspace Shikaze}
\thanks{Present address: Japan Atomic Energy Agency (JAEA), Tokai, Ibaraki, Japan}
\affiliation{Kobe University, Kobe, Hyogo 657-8501, Japan.}
\author{R.\thinspace Shinoda}
\affiliation{The University of Tokyo, Bunkyo, Tokyo 113-0033 Japan.}
\author{R.\thinspace E.\thinspace Streitmatter}
\affiliation{NASA/Goddard Space Flight Center, Greenbelt, MD 20771, USA.}
\author{J.\thinspace Suzuki}
\affiliation{High Energy Accelerator Research Organization (KEK), Tsukuba, Ibaraki 305-0801, Japan.}
\author{Y.\thinspace Takasugi}
\affiliation{Kobe University, Kobe, Hyogo 657-8501, Japan.}
\author{K.\thinspace Takeuchi}
\affiliation{Kobe University, Kobe, Hyogo 657-8501, Japan.}
\author{K.\thinspace Tanaka}
\affiliation{High Energy Accelerator Research Organization (KEK), Tsukuba, Ibaraki 305-0801, Japan.}
\author{N.\thinspace Thakur}
\affiliation{University of Denver, Denver, CO 80208, USA.}
\author{T.\thinspace Yamagami}
\affiliation{Institute of Space and Astronautical Science, Japan Aerospace Exploration Agency (ISAS/JAXA), Sagamihara, Kanagawa 252-5210, Japan.}
\author{A.\thinspace Yamamoto}
\affiliation{High Energy Accelerator Research Organization (KEK), Tsukuba, Ibaraki 305-0801, Japan.}
\affiliation{The University of Tokyo, Bunkyo, Tokyo 113-0033 Japan.}
\author{T.\thinspace Yoshida}
\affiliation{Institute of Space and Astronautical Science, Japan Aerospace Exploration Agency (ISAS/JAXA), Sagamihara, Kanagawa 252-5210, Japan.}
\author{K.\thinspace Yoshimura}
\affiliation{High Energy Accelerator Research Organization (KEK), Tsukuba, Ibaraki 305-0801, Japan.}

\date{\today}

\begin{abstract}
In two long-duration balloon flights over Antarctica, the BESS-Polar collaboration has searched for antihelium in the cosmic radiation with higher sensitivity than any reported investigation. BESS-Polar I flew in 2004, observing for 8.5 days. BESS-Polar II flew in 2007-2008, observing for $24.5$ days. No antihelium candidate was found in BESS-Polar I data among $ 8.4 \times 10^6$ $|Z|=2$ nuclei from $1.0$ to $20$ GV or in BESS-Polar II data among $4.0 \times 10^7$ $|Z|=2$ nuclei from $1.0$ to $14$ GV. Assuming antihelium to have the same spectral shape as helium, a 95\% confidence upper limit of $6.9 \times 10^{-8}$ was determined by combining all the BESS data, including the two BESS-Polar flights. With no assumed antihelium spectrum and a weighted average of the lowest antihelium efficiencies from $1.6$ to $14$ GV, an upper limit of $1.0 \times 10^{-7}$ was determined for the combined BESS-Polar data. These are the most stringent limits obtained to date.
\end{abstract}


\maketitle

\section{Introduction}
The existence of antiparticles was predicted by Dirac \cite{1928RSPSA.117..610D} and confirmed by Anderson through the discovery of the positron, antiparticle of the electron, in the cosmic radiation \cite{1933PhRv...43..491A}. This was followed by experimental confirmation of the existence of antiprotons in the laboratory \cite{1955PhRv..100..947C} and in the cosmic radiation \cite{1979PhRvL..43.1196G,1979ICRC....1..330B}. The production of antinuclei in the laboratory with $|Z|=2$ has now been confirmed \cite{2011Natur.473..353T}. However, in spite of many efforts to find them, there is no evidence that antinuclei with $|Z| \geq 2$ exist in the cosmic radiation \cite{2008AdSpR..42..450S}, or by implication in the universe at large.

The apparent asymmetry of particles and antiparticles is one of the fundamental problems in cosmology. This was probably caused by symmetry-breaking between particles and antiparticles just after the Big Bang, with cosmological antiparticles vanishing at an early stage of the universe. However, local symmetry breaking is not excluded and antimatter domains could remain. Gamma-ray searches for annihilation signatures have set a limit on how near Earth these could be \cite{1998ApJ...495..539C}. However, discovery of complex ($|Z| \ge 2$) cosmic-ray antinuclei would indicate that antimatter domains still exist.  

The BESS collaboration has searched for antinuclei in the cosmic radiation since 1993, with eight conventional one-day balloon flights and two long-duration Antarctic flights.

\section{The BESS-Polar Spectrometer}

\begin{figure*}
  \includegraphics[width=6.8in]{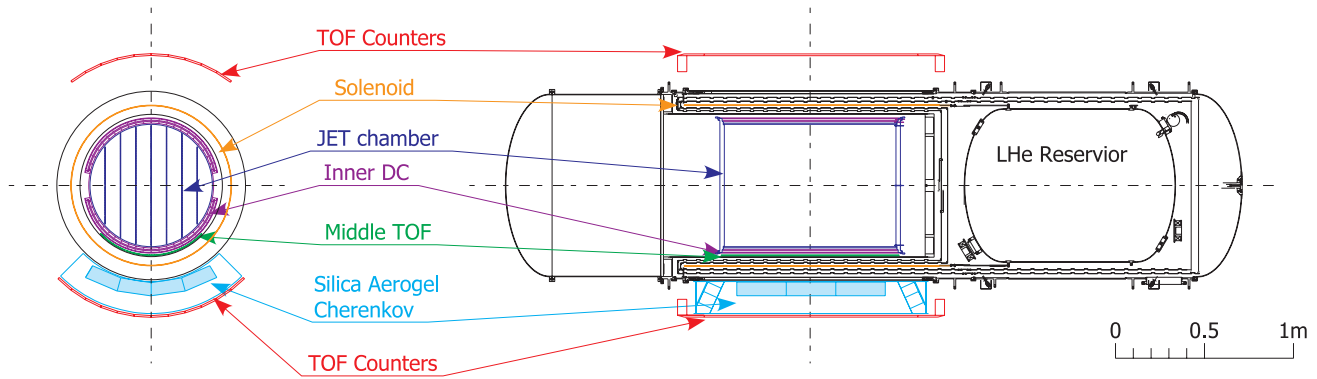}
  \caption{Cross-sectional and side views of the BESS-Polar II Spectrometer.\label{cross_sectional}}
\end{figure*}

The BESS-Polar magnetic-rigidity spectrometer \cite{2002AdSpR..30.1253Y} was developed for precise measurements of cosmic-ray antiprotons to low energies \cite{2011arXiv1107.6000A} and to search for antihelium with great sensitivity. Versions made long-duration balloon flights over Antarctica in 2004 (BESS-Polar I) and 2007-2008 (BESS-Polar II). BESS-Polar is configured to extend measurements down to 100 MeV \cite{2002AdSpR..30.1253Y}. To reduce material encountered by incident particles, no pressure vessel is used and the thickness of the magnet wall is half that of the previous BESS. The time-of-flight (TOF) detectors and aerogel Cherenkov counter (ACC), with their front-end electronics, operated in vacuum. The magnet cryostat was used as the pressure vessel for the central tracker \cite{2004PhLB..594...35H}. The basic spectrometer configuration was the same for BESS-Polar I and BESS-Polar II. For BESS-Polar II, a new magnet with greater liquid helium capacity and improved thermal performance enabled extended observation time.

Figure \ref{cross_sectional} shows schematic cross-sectional and side views of the BESS-Polar II spectrometer. All the detector components are arranged in a cylindrical configuration to maximize geometric acceptance. The TOF scintillators, 10 upper (UTOF) and 12 lower (LTOF), measured incident particle velocities, $\beta = v/c$, with a time resolution of 120 ps and provided independent d$E$/d$x$ measurements. Photomultiplier tubes (PMTs) were coupled to both ends of the scintillator paddles through acrylic light guides. An additional TOF layer (MTOF) was installed between the bottom IDC and warm bore to detect low energy particles that could not penetrate the lower magnet wall. Events were triggered by the UTOF in coincidence with LTOF or MTOF and all were recorded. The ACC was located between the magnet and the LTOF to separate antiproton events from $e^{-}$ and $\mu^{-}$ background. The MTOF and ACC were not used for the antihelium search.

The superconducting solenoid provided a uniform field of 0.8 Tesla for over 11 days continuous operation in BESS-Polar I and over 25 days in BESS-Polar II. Two inner drift chambers (IDCs) and a JET-cell type drift chamber (JET) were located inside the warm bore (0.80 m in diameter and 1.4 m in length). The axial positions of incident particles were initially determined using the UTOF and LTOF. Final axial positions used the JET and IDC. In the bending plane, particle trajectories were fit using up to 52 points, each with 140 $\mu$m resolution. The resulting magnetic-rigidity (R $\equiv$ pc/Ze, momentum divided by electric charge) resolution is 0.4\% at 1 GV, with a maximum detectable rigidity (MDR) of 240 GV. The JET also provided d$E$/d$x$ information. The JET and IDCs used continuously refreshed $CO_2$ gas. 

\section{Flight Conditions}

BESS-Polar I was launched on 13 December 2004 from Williams Field near McMurdo Station. It flew for 8.5 days, recording 900 million cosmic-ray events, and was terminated at the southeast edge of the Ross Ice Shelf. The average altitude was 38.5 km (residual atmosphere of 4.3 $\mathrm{g/cm^2}$). Several PMTs on the TOF that showed extremely high count rates and drew excessive current were turned off, but 66\% of the full geometric acceptance was retained by modifying the trigger algorithm. BESS-Polar II was launched on 22 December 2007. It flew for 29.5 days and observed for 24.5 days at float altitude with the magnet energized, recording 4.7 billion events. Full geometric acceptance was maintained during the entire flight, although two TOF PMTs were turned off due to an HV control issue. After one day, full JET chamber HV could not applied and the gas pressure was adjusted to compensate. The position resolution of the JET chamber was maintained, using HV-dependent calibration over short time intervals, and overall tracking performance was comparable to BESS-Polar I.

\section{Data Analysis}

To eliminate events in which more than one particle passed through the spectrometer and particles interacting in the instrument, events with a single good track were chosen. Only one track was allowed in the drift chamber, and one hit each in the UTOF and LTOF. Next, track quality selections were applied, including hit data consistency between TOF and drift chambers, small $\chi^{2}$ in trajectory fitting, and fiducial cuts. None of these depend on the sign of the particle charge.

\begin{figure}
  \includegraphics[width=3.4in]{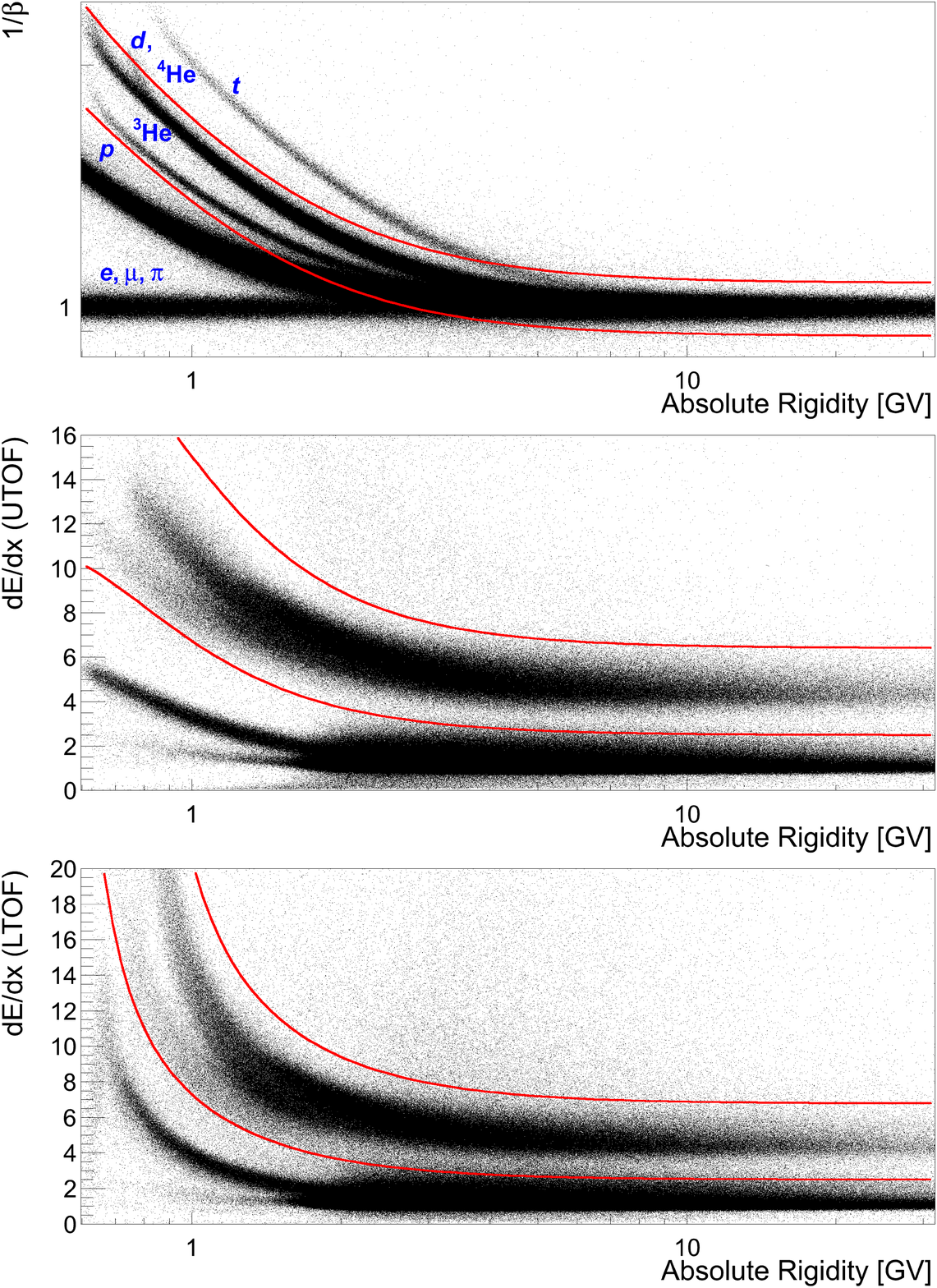}
  \caption{Selection of helium (antihelium) in BESS-Polar II. The upper panel shows $\beta^{-1}$ vs absolute rigidity. The lower two panels show d$E$/d$x$ from the TOF vs absolute rigidity. The $|Z|=2$ particles are between the lines.}
  \label{bess_select}
\end{figure}

Helium (antihelium) nuclei were identified by their absolute charge, $|Z|$ (d$E$/d$x$), and mass, $M$, determined from rigidity $R$, velocity $\beta$ by:
\begin{equation} 
  \label{eq:mass} 
  \displaystyle M^2 = R^2Z^2(\frac{1}{\beta^2}-1). 
\end{equation}
$\beta$, d$E$/d$x$ and $R$ were measured by the TOF and the drift chamber. $1/\beta$ and d$E$/d$x$ band cuts were used to select helium (antihelium), as illustrated for the TOF in Figure \ref{bess_select}. A similar cut was applied to d$E$/d$x$ measured by the drift chamber.

\section{Results}

\begin{figure}
  \includegraphics[width=3.2in]{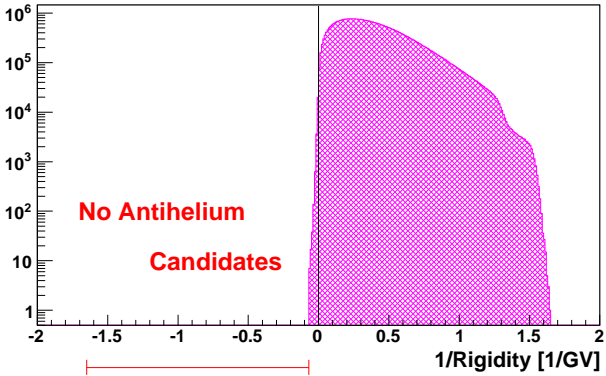}
  \caption{$R^{-1}$ distribution of $|Z|=2$ events for BESS-Polar II.}
  \label{bess_rinv}
\end{figure}

Figure \ref{bess_rinv} shows the $R^{-1}$ distribution of the BESS-Polar II $|Z|=2$ data with all selections applied. The rigidity range for the antihelium search is bounded by rapidly decreasing efficiency at the low end and by the inability of the spectrometer to identify the charge-sign of helium (spill-over) at the high end. No antihelium candidates were found in the rigidity range $1.0$ to $20$ GV, among $8.4 \times 10^{6}$ $|Z|=2$ nuclei identified by BESS-Polar I, or in the rigidity range $1.0$ to $14$ GV, among $4.0 \times 10^{7}$ $|Z|=2$ nuclei identified by BESS-Polar II. Only upper limits to the abundance ratio of antihelium/helium at the top of the atmosphere (TOA) can be determined. 

If antihelium had been observed, the ratio corrected to TOA would have been:
\begin{eqnarray}
  \label{eq:limit}
  & & {} \displaystyle R_{\overline{\mathrm{He}} / \mathrm{He}}  =  \nonumber \\
  & & {} \frac{ \displaystyle \int N_{Obs,{\overline{\mathrm{He}}}} / (S\Omega \times {\overline{\eta}} \times {\overline{\epsilon}}_{sngl} \times {\overline{\epsilon}}_{dE/dx} \times {\overline{\epsilon}}_{\beta} \times {\overline{\epsilon}}_{DQ}) dE}{\displaystyle \int N_{Obs,\mathrm{He}} / (S\Omega \times \eta \times \epsilon_{sngl} \times \epsilon_{dE/dx} \times \epsilon_{\beta} \times \epsilon_{DQ}) dE }, \nonumber \\
\end{eqnarray}
where $N_{Obs, \mathrm{He} (\hebar)}$ is the number of observed He ($\hebar$) events, S$\Omega$ is geometric acceptance, $\eta$ ($\overline{\eta}$) is the survival probability of He ($\hebar$) traversing the atmosphere, $\epsilon_{sngl}$ ($\overline{\epsilon}_{sngl}$) is the single track efficiency for He ($\hebar$), $\epsilon_{dE/dx}$ ($\overline{\epsilon}_{dE/dx}$) is the d$E$/d$x$ selection efficiency for He ($\hebar$), $\epsilon_{\beta}$ ($\overline{\epsilon}_{\beta}$) is the $\beta$ selection efficiency for He ($\hebar$), and $\epsilon_{DQ}$ ($\overline{\epsilon}_{DQ}$) is the data quality selection efficiency for He ($\hebar$).

In order to calculate an upper limit, the energy dependent efficiencies for antihelium must be determined. We calculate upper limits under two different assumptions.

\begin{figure}
  \includegraphics[width=3.4in]{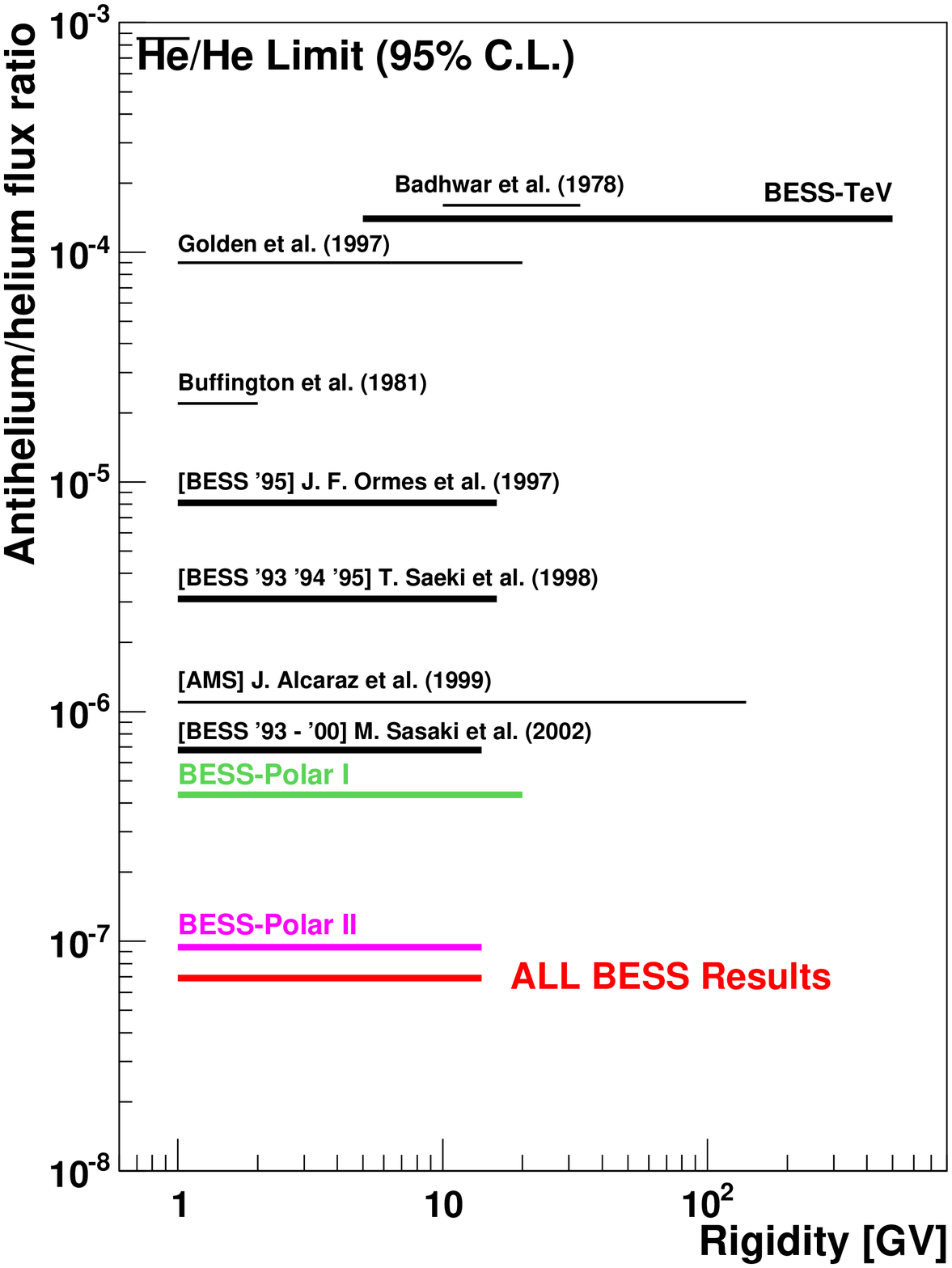}
  \caption{The new upper limits of $\hebar$/He with previous experimental results.(\cite{2008AdSpR..42..450S},\cite{1978Natur.274..137B},\cite{1997ApJ...479..992G},\cite{1981ApJ...248.1179B},\cite{1997ApJ...482L.187O},\cite{1998PhLB..422..319S},\cite{1999PhLB..461..387A},\cite{2002NuPhS.113..202S})}
  \label{upper_limit}
\end{figure}

{\it 1) Same energy spectrum for $\overline{He}$ as for He:}\\
If the hypothetical energy spectrum of antihelium is assumed to be the same as the energy spectrum of helium, Equation \ref{eq:limit} simplifies to:
\begin{eqnarray}
  \label{eq:limit2}
  \displaystyle R_{\overline{\mathrm{He}}/\mathrm{He}} <
  \displaystyle \frac{ 3.1 }{\displaystyle  \int N_{Obs,\mathrm{He}}
    \times {\overline{\eta}} \times {\overline{\epsilon}}_{sngl}
    / ( \eta \times \epsilon_{sngl} ) \mathrm{d}E},
\end{eqnarray}
where 3.1 is the maximum number of hypothetical antihelium nuclei consistent at 95\% confidence with a null detection and no background \cite{1998PhRvD..57.3873F}. $\eta$ $(\overline{\eta})$ and $\epsilon_{sngl}$ $(\overline{\epsilon}_{sngl})$ are determined using a Monte Carlo simulation with GEANT3/GHEISHA. The BESS-Polar I data give an upper limit for $R_{\hebar/\mathrm{He}}$ of $4.4 \times 10^{- 7}$ from $1.0$ to $20$ GV, and the BESS-Polar II data give $9.4 \times 10^{-8}$ from $1.0$ to $14$ GV. Combining the null detections in all BESS flights by summing their Equation \ref{eq:limit2} denominators gives an upper limit of $6.9 \times 10^{-8}$ from $1.0$ to $14$ GV. This is the most stringent upper limit to date. The new limits are shown in Figure \ref{upper_limit} compared with previous results. 

{\it 2) Most conservative limit:}\\
The most conservative upper limit is obtained by applying the lowest overall antihelium efficiency within the search range to any hypothetical $\hebar$. Because S$\Omega$ is nearly constant over the search range, Equation \ref{eq:limit} then simplifies to:
\begin{eqnarray}
  \label{eq:limit3}
  \displaystyle R_{\overline{\mathrm{He}} / \mathrm{He}}  < \frac{ \displaystyle 3.1~ /~ [ {\overline{\eta}} \times {\overline{\epsilon}}_{sngl} \times {\overline{\epsilon}}_{dE/dx} \times {\overline{\epsilon}}_{\beta} \times {\overline{\epsilon}}_{DQ} ]_{MIN}}{\displaystyle \int N_{Obs,\mathrm{He}} / (\eta \times \epsilon_{sngl} \times \epsilon_{dE/dx} \times \epsilon_{\beta} \times \epsilon_{DQ}) dE }, \nonumber \\
\end{eqnarray}
The calculated overall antihelium efficiencies were flat for most of the rigidities searched above, but decreased at lower rigidities due to annihilation. The ranges searched here were set to simultaneously optimize efficiencies and statistics. The resulting most conservative upper limits are $5.3 \times 10^{-7}$ from $1.5$ to $20$ GV for BESS-Polar I and $1.2 \times 10^{-7}$ from $1.6$ to $14$ GV for BESS-Polar II, only about 25\% higher than the corresponding limits calculated above. Data from the BESS-Polar flights were combined by summing the number of helium and using a weighted average of the antihelium efficiencies, giving a conservative upper limit of $1.0 \times 10^{-7}$ from $1.6$ to $14$ GV. For the present work, earlier BESS flights were not reanalyzed under this assumption.

The BESS-Polar collaboration has established the most stringent limits to date on the possible presence of antihelium in the cosmic radiation.

\begin{acknowledgments}
The authors thank NASA Headquarters, ISAS/JAXA, and KEK for continuous support and encouragement in this United States-Japan cooperative project. Sincere thanks are expressed to the NASA Balloon Program Office at GSFC/WFF, the NASA Columbia Scientific Balloon Facility, the National Science Foundation, and Raytheon Polar Services for their professional support of the Antarctic flights. BESS-Polar is supported in Japan by MEXT grants KAKENHI (13001004; 18104006), and in the U.S. by NASA (10-APRA10-0160; NNX10AC48G).
\end{acknowledgments}

\bibliography{2012jan13}

\end{document}